\documentclass[aps,prb,footinbib,twocolumn,floatfix,longbibliography,superscriptaddress,10pt]{revtex4-2}
\usepackage{stix}
\usepackage{graphicx}
\usepackage{xcolor}
\usepackage{comment}
\usepackage{amsmath}
\usepackage{bm}
\usepackage{mathtools}
\definecolor{webblue}{HTML}{2D3092}
\usepackage[colorlinks, urlcolor=webblue, citecolor=webblue, linkcolor=webblue]{hyperref}
\usepackage[capitalize]{cleveref}
\usepackage{tikz}
\usepackage{chemformula}
\usepackage{placeins}
\usepackage{sidecap}

\sidecaptionvpos{figure}{c}
\sidecaptionvpos{table}{c}

\definecolor{colorhhy}{rgb}{0.9, 0.17, 0.31}
\definecolor{colorlmr}{rgb}{0.1, 0.2, 0.7}

\newcommand{\bv}[1]{\boldsymbol{#1}}
\newcommand{\prlparagraph}[1]{{\itshape{}#1{}.}---}

\newcommand{\titlePaper}{Extended {\itshape s}-wave superconductivity in {\itshape M}-point twisted bilayer \ch{SnSe2}}

\newcommand{\paperAuthors}{
\author{Lennart Klebl}
\email{lennart.klebl@uni-wuerzburg.de}
\affiliation{Institute for Theoretical Physics and Astrophysics and Würzburg-Dresden Cluster of Excellence ctd.qmat, University of Würzburg, 97074 Würzburg, Germany}

\author{Ammon Fischer}
\affiliation{Max Planck Institute for the Structure and Dynamics of Matter, Luruper Chaussee 149, 22761
Hamburg, Germany}
\affiliation{Center for Computational Quantum Physics (CCQ), The Flatiron Institute, New York, New York 10010, USA}

\author{Salahudin V.~Smailagi\'c}
\affiliation{Institute for Theoretical Physics and Astrophysics and Würzburg-Dresden Cluster of Excellence ctd.qmat, University of Würzburg, 97074 Würzburg, Germany}
\affiliation{Leinweber Institute for Theoretical Physics, Stanford University, Stanford, CA 94305, USA}

\author{Ming-Rui Li}
\affiliation{Institute for Advanced Study, Tsinghua University, Beijing 100084, China}
\affiliation{Department of Physics, Princeton University, Princeton, NJ 08544, USA}

\author{Henning Schlömer}
\affiliation{ITAMP, Harvard-Smithsonian Center for Astrophysics, Cambridge, MA 02138, USA}
\affiliation{Department of Physics, Harvard University, Cambridge MA 02138, USA}

\author{Haoyu Hu}
\affiliation{Department of Physics, Princeton University, Princeton, NJ 08544, USA}
\affiliation{Department of Physics, University of Science and Technology of China, Hefei, Anhui 230026, China}

\author{B.~Andrei Bernevig}
\affiliation{Department of Physics, Princeton University, Princeton, NJ 08544, USA}
\affiliation{Donostia International Physics Center (DIPC), Paseo Manuel de Lardizábal. 20018, San Sebastián, Spain}
\affiliation{IKERBASQUE, Basque Foundation for Science, 48013 Bilbao, Spain}

\author{Dante M. Kennes}
\affiliation{Max Planck Institute for the Structure and Dynamics of Matter, Luruper Chaussee 149, 22761 Hamburg, Germany} 
\affiliation{Institut für Theorie der Statistischen Physik, RWTH Aachen, 52056 Aachen, Germany and JARA -- Fundamentals of Future Information Technology}

\author{Ronny Thomale}
\affiliation{Institute for Theoretical Physics and Astrophysics and Würzburg-Dresden Cluster of Excellence ctd.qmat, University of Würzburg, 97074 Würzburg, Germany}
}

\newcommand{\supplement}[1]{%
  \onecolumngrid%
  \clearpage%
  \title{#1}%
  \maketitle%
  \setcounter{equation}{0}%
  \setcounter{figure}{0}%
  \setcounter{table}{0}%
  \setcounter{page}{1}%
  \makeatletter%
  \renewcommand{\thesection}{S\arabic{section}}%
  \renewcommand{\thesubsection}{\Alph{subsection}}%
  \renewcommand{\theequation}{S\arabic{equation}}%
  \renewcommand{\thefigure}{S\arabic{figure}}%
  \renewcommand{\thetable}{S\Roman{table}}%
  \renewcommand{\thepage}{S\arabic{page}}%
  \numberwithin{figure}{section}%
  \numberwithin{table}{section}%
  \numberwithin{equation}{section}%
  \makeatother%
  \onecolumngrid%
}
\makeatletter
\def\maketitle{
\@author@finish
\title@column\titleblock@produce
\suppressfloats[t]}
\makeatother

\let\oldcite\cite
\renewcommand{\cite}[1]{\if\relax\detokenize{#1}\relax\textbf{\color{red}[?]}\else\oldcite{#1}\fi}

\begin{document}
\title{\titlePaper}
\paperAuthors

\begin{abstract}
We investigate the emergence of electronic order and unconventional superconductivity in $M$-valley moiré materials. 
Starting from a first-principles Wannier model of AB-stacked twisted \ch{SnSe2}, we tackle the (gate-screened) long-ranged
Coulomb interaction with functional renormalization group simulations resolving the momentum structure and energy scales of the leading Fermi surface instabilities.
Upon doping an antiferromagnetic stripe state 
at half-filling
($\nu=3$ electrons per moir\'e unit cell)
of the moiré flat bands, magnetic order gives way to unconventional superconductivity mediated by valley-selective spin fluctuations:
Large hole doping ($\nu\approx1$) leads to weak-coupling superconductors with various pairing symmetries, while slight electron- and hole-doping ($\nu\approx2,4$)
stabilizes a spin-singlet, extended $s$-wave state that benefits from scattering between virtual particle and hole states that are detuned from the Fermi level.
These findings establish $M$-point moiré materials as a quantum simulation platform with phenomenological parallels to the class of iron pnictide superconductors.
\end{abstract}
\maketitle

\prlparagraph{Introduction}%
Since the discovery of superconductivity in moir\'e materials~\cite{cao2018unconventional}, the long-standing effort to understand the microscopic mechanisms of unconventional electronic order in high-$T_c$ superconductors---including cuprates~\cite{bednorz1986possible}, heavy-fermion materials~\cite{steglich1979superconductivity}, and iron pnictides~\cite{kamihara2008ironbased}---has been revitalized by the ability to realize analogous quantum phases in tunable van der Waals heterostructures~\cite{kennes2021moire}.
Superconductivity is now firmly established across diverse platforms, including twisted bilayer graphene~\cite{cao2018unconventional, yankowitz2019tuning, lu2019superconductors, stepanov2020untying, saito2020independent, devries2021gatedefined, oh2021evidence, tian2023evidence, dibattista2022revealing, calugaru2022spectroscopy}, twisted multilayer graphene~\cite{chen2019signatures, park2021tunable, hao2021electric, cao2021paulilimit, liu2022isospin, kim2022evidence, zhang2022promotion}, rhombohedral multilayer graphene~\cite{zhou2021superconductivity, zhou2022isospin, zhang2023enhanced, han2025signatures, choi2025superconductivity}, and twisted transition metal dichalcogenides (TMDs) such as \ch{WSe2}~\cite{xia2025superconductivity, guo2025superconductivity, xia2026bandwidthtuned, guo2025angle} and \ch{MoTe2}~\cite{xu2025signatures,sun2026twist}.
A common feature of some of these moir\'e superconductors is that their low-energy electronic states originate from the two inequivalent $K^{\eta}$-points of the monolayer hexagonal Brillouin zone (BZ). Consequently, the low-energy theory inherits an emergent $U^{\eta}(1)$ symmetry; however, because time-reversal symmetry relates the two valleys ($\mathcal{T} K^{\eta} = K^{\bar \eta}$), it is not preserved within an individual valley.
In the context of BCS-like superconductivity, this symmetry structure favors intervalley coherent order parameters, $\Delta \sim \langle c_{\bv k \eta} c_{-\bv k \bar \eta} \rangle$, where Cooper pairs are formed by electrons from different valleys.
While the pairing mechanism and order parameter symmetry in graphene systems remain subjects of debate~\cite{stepanov2020untying, saito2020independent, liu2021tuning, chen2024strong, wang2024molecular}, the situation in $K$-valley TMDs like twisted \ch{WSe2}~\cite{xia2025superconductivity, guo2025superconductivity, xia2026bandwidthtuned, guo2025angle} is more definitive.
Accumulating evidence points toward a weak-to-moderate coupling scenario~\cite{guo2025angle}, for which theory predicts an intervalley coherent superconducting order parameter driven by spin-fluctuation exchange~\cite{klebl2023competition, fischer2025theory,  chubukov2025quantum, qin2025topological} across the experimentally accessible range of twist angles.

In contrast, moir\'e materials based on hexagonal monolayers with low-energy valleys at the three inequivalent $M$-points of the Brillouin zone~\cite{calugaru2025moire, lei2025moire, bao2025anisotropic, ingham2025moire}
feature three time-reversal-invariant valleys, $\mathcal{T}M^{\eta} = M^{\eta}$.
These valleys are related by threefold rotational symmetry, $C_{3z}M^{\eta} = M^{\mathrm{mod}(\eta+1,3)}$, and inherit an approximate spin $SU(2)$ symmetry from the monolayer, thereby offering new avenues for engineering unconventional superconductivity in moir\'e heterostructures.
$M$-valley moir\'e materials can be realized by twisting monolayers of group-IV trigonal transition metal dichalcogenides (1T-\ch{MX2}, where M\,=\,Zr,\,Hf,\,Sn and X\,=\,S,\,Se)~\cite{calugaru2025moire, lei2025moire, jiang2024twodimensional, xu2026organizing}.
Their low-energy theory is characterized by an emergent momentum-space non-symmorphic symmetry~\cite{calugaru2025moire} (which in \ch{SnSe2} gives a one-dimensionality opposite to the one suggested by the mass tensor~\cite{kariyado2019flat, fujimoto2022perfect, kariyado2023twisted}) effectively realizing a six-flavor $U(2)\otimes U(2) \otimes U(2)$ Hubbard model in the presence of long-range Coulomb interactions.
Previous theoretical studies~\cite{li2025emergent, beule2025role, park2026intervalley} have explored the complex landscape of magnetic and charge order in $M$-valley AA- and AB-stacked twisted \ch{SnSe2} using mean-field treatments of correlations, starting from either effective Wannier models~\cite{li2025emergent} or continuum Hamiltonians~\cite{beule2025role, park2026intervalley}.
While the low-energy physics of AA-stacked \ch{SnSe2} is quasi-one-dimensional within each valley, acquiring a two-dimensional character only through longer-ranged intervalley (density-density) interactions, the approximate Kagome lattice formed by the Wannier centers in AB-stacked \ch{SnSe2} exhibits a more pronounced two-dimensional character already at the kinetic level, resulting in a less correlated regime.
The bandwidth of the flat bands can also be tuned by increasing the twist angle. 

In this Letter, we address the nature of superconducting order in $M$-valley moir\'e materials, taking AB-stacked twisted \ch{SnSe2} (\ch{tSnSe2}) as a representative example.
This system is particularly suitable for such an investigation because its Hamiltonian possesses a more pronounced two-dimensional character and likely resides in the weak-to-intermediate coupling regime~\cite{li2025emergent}.
We demonstrate that upon slight electron- and hole-doping ($\nu\approx2,4$)
the antiferromagnetic (AFM) state at half-filling ($\nu = 3$), AB-stacked \ch{tSnSe2} hosts an intravalley, extended $s$-wave superconducting state driven by valley-%
selective spin fluctuations. Under large
hole-doping ($\nu\approx 1$), the Fermi surface topology instead favors extended $s$-, $p$-, or $d$-wave superconductivity; all in close energetic proximity.
Interestingly, the superconducting order identified in AB-stacked \ch{tSnSe2} exhibits striking parallels to the class of iron pnictide superconductors~\cite{kamihara2008ironbased, stewart2011superconductivity, si2016hightemperature}. 
This analogy refers to the stabilization of an extended $s$-wave order parameter~\cite{mazin2008unconventional, kuroki2009pnictogen, hirschfeld2011gap, chubukov2012pairing, thomale2011mechanism, platt2011superconducting} via virtual spin-fluctuation exchange between Fermi pockets slightly detuned from the Fermi level~\cite{xiang2012hightemperature}, as well as the emergence of higher angular momentum pairing upon hole doping~\cite{thomale2009functional}.
We therefore propose $M$-point moir\'e materials as a complementary platform for investigating the microscopic mechanisms of high-$T_c/T_F$ superconductivity, extending the field beyond the heavy-fermion physics of twisted bilayer graphene~\cite{song2022magicangle, datta2023heavy, rai2024dynamical}
and the AFM-mediated pairing observed in $K$-valley \ch{tWSe2}~\cite{xia2026bandwidthtuned, guo2025angle}.

\begin{figure}
    \centering
    \includegraphics{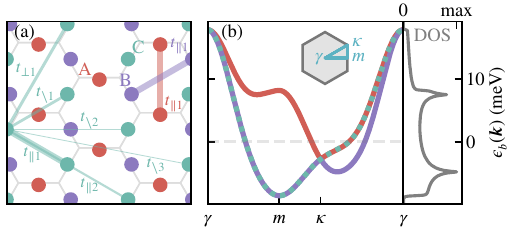}
    \caption{Wannier centers, hopping parameters~(a), and band structure~(b) of $\theta=6.01^\circ$ twisted bilayer \ch{SnSe2} in AB stacking configuration.
    We visualize the hopping parameters for the $\mathrm C$ valley as green lines with thickness encoding strength, see \cref{tab:wannparam} for numerical values. The hoppings for $\eta = \mathrm{A, B}$ follow from $C_{3z}$ rotation, as indicated by the dominant hopping $t_{\parallel1}$. The Wannier centers (dots) are located approximately on the three Kagome sites in the Wigner-Seitz cell (light gray hexagons)
    The bands in panel~(b) are colored according to their valley content ($\eta=\mathrm{C}$ dotted for visibility)
    and shown along the irreducible path indicated in the inset.
    The dashed gray line indicates the position of the Fermi level for half filling ($\nu=3$).
    The density of states (DOS) reveals the presence of two Van Hove singularities at approximately equal distance from half filling.
    }
    \label{fig:model}
\end{figure}
\prlparagraph{Model}%
We construct an effective low-energy tight-binding model for AB-stacked \ch{tSnSe2} following the procedure outlined in Ref.~\cite{li2025emergent}. This approach is based on a continuum model derived from \emph{ab initio} simulations of twisted \ch{SnSe2} bilayers, followed by the Wannierization of the three lowest conduction bands~\cite{kariyado2023twisted, lei2025moire, calugaru2025moire, bao2025anisotropic}. Each of the three bands in the continuum model can be assigned a well-defined valley quantum number $\eta \in \{\mathrm{A}, \mathrm{B}, \mathrm{C} \}$ due to the emergent $U^{\eta}(1)$ symmetry. Furthermore, these bands are topologically trivial, allowing for the construction of exponentially localized Wannier functions. The resulting lattice model is displayed in \cref{fig:model}(a), where the three Wannier orbitals corresponding to the three distinct valleys ($M$-points) form an approximate Kagome lattice. The non-interacting Hamiltonian $H_0$ is diagonal in the valley index $\eta$ and reads
\begin{equation}
    \label{eq:hkin}
    H_0 = \sum_{\bv R\bv \Delta,\eta, \sigma} t_{\bv\Delta,\eta}^{\vphantom{\dagger}}  c^\dagger_{\bv R\vphantom{'},\eta,\sigma} c^{\vphantom{\dagger}}_{\bv R+\bv \Delta,\eta,\sigma} \,,
\end{equation}
where $c^{(\dagger)}_{\bv R,\eta,\sigma}$ annihilates (creates) an electron in valley $\eta$ with spin $\sigma$ on site $\bv R$. 
The multi-orbital Wannier model of AB-stacked \ch{tSnSe2} is characterized by the symmetry group $\mathcal{G}_0 = D_{3} \otimes \mathcal{T} \otimes U^{\eta}(2)$. The moir\'e point group $D_{3}$ is generated by the threefold rotation $C_{3z}$ and twofold rotation $C_{2y}$; the latter acts as a mirror symmetry $\mathcal{M}_{\eta}$ along the valley's preferred hopping direction when projected to two dimensions (space group $P321$, \#150). Each valley further retains time-reversal $\mathcal{T}$ and spin-$SU(2)$ symmetry, leading to an effective $U^{\eta}(2)$ symmetry per valley. The conduction band structure of AB-stacked \ch{tSnSe2} is shown in \cref{fig:model}(b) and reveals rather anisotropic hopping parameters $t_{\bv \Delta,\eta}$, with the respective hopping's strength indicated as line thickness for $\eta=\mathrm{C}$ in \cref{fig:model}(a), see \cref{tab:wannparam} for numerical values.
The density of states (DOS) features two Van Hove singularities that are not pinned to high-symmetry momenta. This is a consequence of the fact that $C_{3z}$ rotations map valleys onto one another, rather than constraining the dispersion within a single valley individually.

The interacting part of the Hamiltonian is modeled by a dual-gated Coulomb repulsion of the form $V(\bv q) = 2\pi e^2\tanh(\xi|\bv q|/2)/(\epsilon|\bv q|)$, where $\xi=10\,\mathrm{nm}$ represents the screening length determined by the distance to the experimental gates and $\epsilon$ is the effective dielectric constant. After projection onto the Wannier basis and transformation to real space, we obtain the interaction Hamiltonian
\begin{equation}
    H_I = \hspace{-1.5ex}\sum_{\bv R\bv \Delta,\eta\eta',\sigma\sigma'}\hspace{-1.5ex} V_{\eta,\eta'}(\bv \Delta)\,c^{\dagger}_{\bv R,\eta,\sigma}c^{\vphantom{\dagger}}_{\bv R,\eta,\sigma}\,c^{\dagger}_{\bv R+\bv \Delta,\eta',\sigma'}c^{\vphantom{\dagger}}_{\bv R+\bv \Delta,\eta',\sigma'} \,.
    \label{eq:hint}
\end{equation}
In principle, the projection to the Wannier basis also generates interaction terms that are not of the density-density type. However, given the $s$-wave character of the orbitals and their distinct basis positions, these nontrivial interaction elements are strongly suppressed ($\lesssim 2\%$ of the Hubbard-$U$).
Consequently, we restrict our treatment to the density-density terms in \cref{eq:hint}; see \cref{app:wannier-interaction} and \cref{tab:vertices} for further details.

\prlparagraph{Functional renormalization group analysis}%
Next, we investigate the correlation physics captured by the interacting Wannier model $H=H_0+H_I$, as defined in \cref{eq:hkin,eq:hint}. In experimental settings, the overall interaction scale is often difficult to determine accurately through theoretical calculations alone, primarily due to uncertainties in the dielectric constant $\epsilon$ and screening effects from omitted bands~\cite{profe2025exact}. Consequently, we treat $\epsilon$---which controls the global interaction strength---as a tuning parameter. Since smaller (larger) twist angles typically correspond to larger (smaller) effective interaction scales~\cite{guo2025angle}, we fix the twist angle to $\theta=6.01^\circ$ and explore the phase diagram specifically as a function of the dielectric environment.

To obtain an unbiased characterization of the system's phase diagram in the weak- to intermediate-coupling regime, we employ the functional renormalization group (FRG)~\cite{metzner2012functional, platt2013functional, dupuis2021nonperturbative}. This method accounts for particle-particle and particle-hole fluctuations, as well as their mutual crosstalk, by flowing from the bare action to the fully interacting theory via a regulator $R(\Lambda)$. We adopt the standard approximations of (i)~restricting the FRG flow to the two-particle vertex $\Gamma^{(4)}$ and (ii)~neglecting its frequency dependence~\cite{metzner2012functional, platt2013functional}. Under these conditions, a divergence of $\Gamma^{(4)}$ at a critical scale $\Lambda_{\mathrm{c}}$ signals an instability toward a symmetry-broken phase. The nature of the transition is then identified from the structure of $\Gamma^{(4)}$: the leading diagrammatic channel---particle-particle ($\mathcal{P}$), crossed particle-hole ($\mathcal{C}$), or direct particle-hole ($\mathcal{D}$)---along with its momentum, orbital, and spin composition, uniquely determines the corresponding order parameter.
We employ a sharp frequency cutoff as the regulator for the bare Green's function, $G^\Lambda(i\omega) = R(\Lambda) G^0(i\omega)$ with $R(\Lambda)=\Theta(|i\omega|-\Lambda)$. This choice of $R(\Lambda)$ implies a zero-temperature formalism, where the critical scale $\Lambda_{\mathrm{c}}$ serves as a proxy for the transition temperature $T_{\mathrm{c}}$~\cite{fischer2025theory, guo2025angle}.
To render the FRG flow equations numerically tractable, we utilize the truncated unity (TU) compression scheme for fermionic momenta~\cite{husemann2009efficient, lichtenstein2017highperformance, beyer2022reference, profe2022tu2frg} (see \cref{app:tufrg} for details).
Notably, FRG without interchannel feedback reduces to individual ladder resummation (i.e., random phase approximation) in each of the channels $\mathcal P$, $\mathcal C$, and $\mathcal{D}$~\cite{fischer2024spin}.

\begin{figure}
    \centering
    \includegraphics{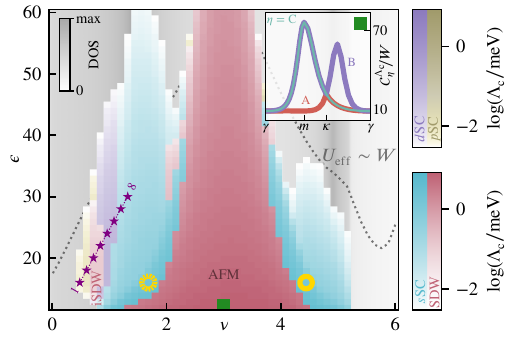}
    \caption{Phase diagram of $\theta=6.01^\circ$ AB-stacked t\ch{SnSe2} obtained by functional renormalization group simulations. We plot the critical scale $\Lambda_\mathrm{c}$, which qualitatively resembles a critical temperature, as a function of electron filling $\nu$ and dielectric constant $\epsilon$.
    We find various regions of distinct phases: An intravalley antiferromagnet (AFM, magenta), an adjacent extended $s$-wave superconductor ($s$SC, blue, yellow circle), $p$- and $d$-wave superconductivity ($p$SC, olive and $d$SC, purple, respectively; purple stars), and incommensurate intravalley magnetism (iSDW, magenta).
    The regions without coloring correspond to Fermi liquid behavior (i.e., no divergence within the FRG) and show the non-interacting density of states (DOS) as background.
    The gray dotted horizontal line indicates the value of $\epsilon$ at which the Hubbard interaction $U_\mathrm{eff}$ is screened to a value smaller than the bandwidth $W$.
    The inset displays the intravalley on-site component of the renormalized vertex in the magnetic channel ($\mathcal C^{\Lambda_\mathrm{c}}_\eta$) along the irreducible path for $\nu=3$ and $\epsilon=12$ (green square, in units of bandwidth $W$).
    The coloring of valleys follows \cref{fig:model}, as does the irreducible path. The peak structure of the vertex corresponds to AFM order along the one-dimensional direction.
    }
    \label{fig:frg-phases}
\end{figure}

\begin{figure}
    \centering
    \includegraphics{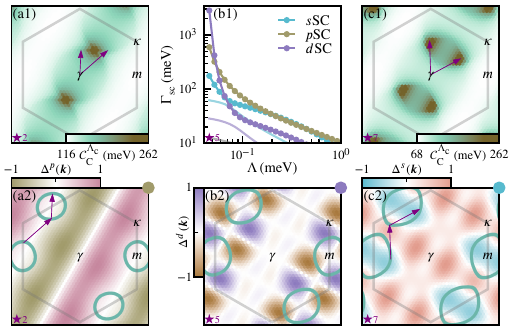}
    \caption{Weak-coupling superconductivity in $\theta=6.01^\circ$ AB-stacked \ch{tSnSe2} (cf.~purple stars in \cref{fig:frg-phases}, counted left-to-right). Panel~(a1) displays the spin fluctuation vertex (at the second purple star in \cref{fig:frg-phases}) from FRG as a function of transfer momentum $\bv q$
    for valley $\eta=\mathrm{C}$, with relevant transfer momenta given by the purple arrows. The resulting superconducting gap amplitude is given in panel~(a2). The leading spin fluctuation vectors (purple) connect parts of the Fermi surface (green circles) where the sign of the superconducting gap $\Delta^p(\bv k)$ changes.
    Panels~(c1,c2) show the same analysis but for the weak-coupling extended $s$-wave case (seventh purple star in \cref{fig:frg-phases}).
    For fillings in between the $s$SC and $p$SC region (fifth purple star in \cref{fig:frg-phases}), the spin fluctuation picture alone is insufficient to predict the $d$SC found in FRG. The flow of the superconducting vertex $\Gamma_\mathrm{sc}$ as a function of RG scale $\Lambda$~(b1) displays close competition of all three states, with $d$SC ultimately dominating~(b2).}
    \label{fig:pwave}
\end{figure}

\begin{figure}
    \centering
    \includegraphics{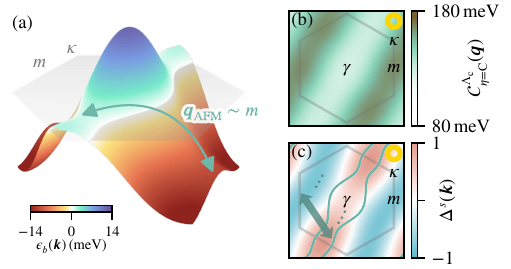}
    \caption{Microscopic pairing mechanism for extended $s$-wave superconductivity in $\theta=6.01^\circ$ AB-stacked \ch{tSnSe2}.
    Panel~(a) illustrates the dominant AFM spin fluctuations' origin starting from the non-interacting band structure (shown as colored three-dimensional surface for valley $\eta=\mathrm{C}$, with the Fermi level being white).
    The green arrow sketches the leading off-shell magnetic scattering, which expresses transfer momenta located in a strip parallel to the Fermi surface (containing $\bv q_\mathrm{AFM}\sim m$);
    panel~(b) shows the $\mathcal C$-channel vertex' structure (for $\nu=4.44$ and $\epsilon=16$; golden circle in \cref{fig:frg-phases}) in the Brillouin zone, where the brown regions contain $\bv q_\mathrm{AFM}\sim m$.
    In panel~(c) we plot the C-valley superconducting order parameter $\Delta^s(\bv k)$ as obtained from FRG. We overlay the Fermi surface in green and indicate the AFM scattering vectors $\bv q_\mathrm{AFM}$ as thick green arrow that can be continued along the dots to retrieve its banded character. The sign change between $\Delta^s(\bv k)$ and $\Delta^s(\bv k + \bv q_\mathrm{AFM})$---which is required for generating (attractive) pairing glue---happens off-shell as well, evidenced by the largely constant gap on the Fermi surface.
    }
    \label{fig:mechanism}
\end{figure}

\prlparagraph{Quantum phases in AB-stacked \ch{tSnSe2}}%
\Cref{fig:frg-phases} displays the FRG phase diagram of $\theta=6.01^\circ$ \ch{tSnSe2} as a function of electron filling $\nu$ and dielectric constant $\epsilon$ (see Ref.~\cite{frg-details}
for technical details). Consistent with previous studies~\cite{li2025emergent, beule2025role}, electronic correlations favor intravalley order across the full filling range; due to the $U^{\eta}(1)$ symmetry, the ground state remains threefold degenerate across the phase diagram. In \cref{fig:frg-phases}, we encode the nature of the leading (intravalley) order as color and the critical scale $\Lambda_\mathrm{c}$ as intensity. Blue, olive, and purple regions correspond to extended $s$-wave ($s$SC), $p$-wave ($p$SC), and $d$-wave ($d$SC) superconductivity, respectively, while magenta indicates a spin density wave; uncolored regions represent a Fermi liquid. The DOS of the non-interacting (i.e., independent on $\epsilon$) model (gray) is provided in the background as a reference.
The center SDW region is of AFM type and emerges around half-filling ($\nu=3$) irrespective of the interaction strength, while a small pocket of incommensurate spin density wave is found at the interface of $p$SC and $d$SC at $\nu\lesssim 1$.
In the inset of \cref{fig:frg-phases} (green square at $\nu=3$, $\epsilon=12$), we plot the transfer momentum $\bv{q}$ dependence of the renormalized vertex in the crossed particle-hole (magnetic) channel $\mathcal C^{\Lambda_\mathrm{c}}(\bv q)$ along the high-symmetry path.
The leading components of the vertex inherit the hoppings' intravalley character, such that we can give $\mathcal C_\eta^{\Lambda_\mathrm{c}}(\bv q)$ a valley index $\eta$ and plot the intravalley vertex for the three valleys $\eta\in\{\mathrm{A,B,C}\}$.
The leading transfer momenta, associated with the peaks of $\mathcal C_\eta^{\Lambda_\mathrm{c}}(\bv q)$, correspond to those $m$-points where the (in general incommensurate) magnetic order
aligns with the dominant hopping direction (cf.~\cref{fig:model}). The momentum-space profile is notably flat along the subdominant hopping direction, as indicated by the secondary peak near $\kappa$ (cf.~\cref{app:banded-vertex}). We attribute the formation of this quasi-one-dimensional AFM order to the fact that, near half-filling, most spectral weight above the Fermi level is connected to that below the Fermi level by a transfer vector $\bv q\approx m$ [cf.~\cref{fig:mechanism}(a)].
This leads to a pronounced peak in the particle-hole susceptibility, which the FRG amplifies to trigger a Stoner-like instability. 
The AFM region is flanked by $s$SC for fillings both above and below $\nu=3$.
Around the two Van Hove fillings ($\nu\approx1$ and $\nu\approx 5$), superconducting domes that extend to large $\epsilon$ (weak coupling) emerge at relatively low critical scales $\Lambda_\mathrm{c} \lesssim 0.1\,\mathrm{meV}$.
In the low filling regime ($\nu\approx 1$),
the Lifshitz transition associated with the lower Van Hove singularity transforms the Fermi surface from lines into small, two-dimensional pockets at the Brillouin zone edges [cf.~\cref{fig:pwave} and \cref{fig:mechanism}(c)].
In the nearly full regime ($\nu\approx 5$), the Van Hove point emerges on the perpendicular $\gamma m$-line (instead of $\gamma\kappa$), cf.~local minimum of the orange $\eta=\mathrm{A}$ band in \cref{fig:model}(b).

\prlparagraph{Microscopic Pairing Mechanism}%
All superconducting instabilities ($s$SC, $p$SC and $d$SC) are found in close proximity to magnetic order (AFM, iSDW), suggesting spin fluctuations as a unified pairing glue in our FRG analysis. More fundamentally, the superconducting phases fall into two regimes distinguished by their critical scale: around the Van Hove fillings, low $\Lambda_\mathrm{c}\lesssim0.1\,\mathrm{meV}$ confines pairing to the Fermi surface, whereas flanking the AFM dome, the larger $\Lambda_\mathrm{c}$ admits pairing from virtual states detuned from the Fermi level.
The system's quasi-one-dimensional character implies that the primary candidates for the superconducting gap are even ($s$-wave) and odd ($p$-wave) symmetries with respect to the parallel mirror $\mathcal M_\eta$, as these are the natural irreps in 1D.
Both are realized in the phase diagram (cf.~\cref{fig:frg-phases}), with the selection determined by the Fermi surface topology and the dominant spin-fluctuation content.

To substantiate the above intuition, we reduce the diagrammatic content of the FRG to explicitly construct effective spin-fluctuation vertices. We employ the intraorbital bilinear approximation (cf.~\cref{app:iobi}), which extends RPA by incorporating screening from the $\mathcal{P}$ and $\mathcal{D}$ channels into the effective Hubbard interaction $U_\mathrm{eff}$, thereby suppressing the critical scale $\Lambda_\mathrm{c}$. By projecting the resulting spin-fluctuation vertex ($\mathcal{C}$-channel) into the superconducting ($\mathcal{P}$) channel, we solve the linearized gap equation,
\begin{multline}
    \label{eq:lingap}
    \lambda\,\Delta^{S/T}_\eta(\bv k) = -\frac12\sum_{\bv k'} \big( \mathcal C_\eta^{\Lambda_\mathrm c}(\bv k+\bv k') \pm \mathcal C_\eta^{\Lambda_\mathrm c}(\bv k-\bv k')\big) \\ \chi^P_{\eta,\Lambda_\mathrm{c}}(\bv k')\, \Delta^{S/T}_\eta(\bv k') \,,
\end{multline}
for the singlet ($S$) or triplet ($T$) intravalley superconducting gap function $\Delta^{S/T}_\eta(\bv k)$, where the eigenvalue $\lambda \to 1$ at the critical temperature (cf.~\cref{app:lingap} for details).

Solving \cref{eq:lingap} reveals that the $p$SC state (first two purple stars in \cref{fig:frg-phases}, counted from left) follows the standard weak-coupling principle: At low critical scales, pairing is governed by a discrete set of fluctuation vectors connecting the nearly isotropic Fermi pockets, and the gap acquires the minimal number of nodes compatible with the leading spin fluctuations---i.e., $\Delta(\bv k)$ changes sign upon translation by the leading (incommensurate) fluctuation vector such that the overall minus sign in \cref{eq:lingap} is canceled. The node positions are thus dictated by the spin-fluctuation content and pinned along the quasi one-dimensional direction [\cref{fig:pwave}(a1,a2)]. At slightly larger hole doping, the $p$SC phase gives way to a narrow $d$SC region. The $d$-wave gap transforms identically to $p$SC under $\mathcal M_\eta$, differing only by an additional node on the Brillouin zone boundary perpendicular to the quasi one-dimensional direction. Since the isotropic pockets barely penalize this node, $d$SC remains only weakly subleading to $p$SC in the pure spin-fluctuation treatment of \cref{eq:lingap}; the interchannel feedback captured by full FRG then overturns this hierarchy in favor of $d$SC [\cref{fig:pwave}(b1,b2)].

Further increasing $\nu$ along the line of purple stars in \cref{fig:frg-phases} drives the system into an $s$SC state [cf.~\cref{fig:pwave}(c1,c2)], where the spin-fluctuation treatment of \cref{eq:lingap} again reproduces the leading instability found in FRG. The simplified $\mathcal C$-channel analysis thus agrees with full FRG on both sides of the $d$SC region, which remains the only regime where interchannel feedback qualitatively alters the pairing symmetry. At larger electron doping ($\nu\approx5$), $s$SC is the only superconducting instability that remains. This can be understood from the distinct Fermi surface geometries near the two Van Hove singularities: while the Lifshitz transition at $\nu\approx1$ produces closed, nearly circular pockets that do not penalize any nodal orientation---enabling the close competition between $s$-, $p$-, and $d$-wave pairing discussed above---the Fermi surface near $\nu\approx5$ retains much of its line-like character [cf.~\cref{fig:mechanism}(c)] with elongated instead of circular pockets (cf.~\cref{fig:supp-fs}). This renders higher-harmonic pairing states with nodes in the two-dimensional direction energetically costly and leaves the extended $s$-wave state as preferred instability.

While the $s$SC states near the AFM dome ($\nu\approx2,4$; golden circles in \cref{fig:frg-phases}) are also captured by the spin-fluctuation analysis, their comparatively large critical scale $\Lambda_\mathrm{c}$ (and hence higher critical temperature) warrants a closer examination of their microscopic origin.
In contrast to the low-$\Lambda_\mathrm{c}$ regime, where the particle-particle bubble $\chi^P$ in \cref{eq:lingap} sharply projects the pairing interaction onto the Fermi surface [cf.~\cref{fig:suppchi}(a)], the large critical scale near the AFM dome renders the algebraic decay of $\chi^P$ away from the Fermi level decisive: off-shell states contribute significantly to the pairing [cf.~\cref{fig:suppchi}(b)].
\Cref{fig:mechanism} illustrates the microscopic origin of this off-shell extended $s$-wave pairing. The leading spin fluctuations are those that stabilize the dominant AFM state near $\nu=3$; they originate from particle-hole excitations between virtual states detuned from the Fermi level (green arrow in \cref{fig:mechanism}(a), cf.~\cref{app:banded-vertex}). At scales $\Lambda$ comparable to the bandwidth $W$,
these states
determine
the particle-hole susceptibility because their scattering vector $\bv q_\mathrm{AFM} \sim m$ connects regions with a high density of states. 
When overlaying the banded AFM spin-fluctuation vectors
[cf.~\cref{fig:mechanism}(b)] with the Fermi surface in \cref{fig:mechanism}(c), we find that they connect regions on the Fermi surface (where $\Delta^s< 0$) to regions \emph{away} from the Fermi surface where the gap changes sign ($\Delta^s > 0$; thick green arrow). The total attraction in the pairing channel is enhanced by the fact that the AFM vertex remains large for all transfer vectors parallel to the Fermi surface, which are summed over in the linearized gap equation \cref{eq:lingap}. In this system, the competition between (i)~the increased pairing strength afforded by the quasi-one-dimensional geometry and (ii)~the suppression of pairing by the algebraic decay of the off-shell particle-particle bubble is resolved in favor of these virtual processes. Ultimately, this results in an extended $s$-wave superconducting gap with negligible nodes on the Fermi surface, as visualized in \cref{fig:mechanism}.

\prlparagraph{Summary \& Outlook}%
Our work provides an unbiased characterization of the weak- to intermediate-coupling phase diagram of $M$-valley moir\'e materials, specifically AB-stacked $6.01^\circ$-twisted \ch{SnSe2}. Distinct from other twisted material platforms, the intravalley spin-$SU(2)$ and time-reversal $\mathcal{T}$ symmetries strongly promote intravalley physics, while the fermiology in each valley transitions from a quasi-one-dimensional (banded) to a two-dimensional (pocket) character upon doping.
Our study demonstrates that the quasi-one-dimensional character of AB-stacked \ch{tSnSe2} is essential,
as line-shaped Fermi surfaces allow for the dominance of high-energy AFM fluctuations with a banded character. These fluctuations drive unconventional superconductivity at relatively large energy scales (i.e., relatively large $T_c/T_F$), overcoming the algebraic energetic penalty for off-shell sign changes in the order parameter and resulting in the aforementioned \emph{quasi-nodeless} state. We expect this mechanism to carry over to other material platforms situated in the dimensionality crossover regime~\cite{cho2013band}.

Since the three valleys are kinetically decoupled, the superconducting state in \ch{tSnSe2} can be interpreted as three independent, highly anisotropic, and weakly Josephson-coupled superconductors. Exploring the applications of this peculiar superconducting phase, beginning with a mean-field treatment of our microscopic model, remains a compelling direction for future study. Furthermore, it is essential to benchmark our FRG approach—which originates from the two-dimensional limit yet predicts quasi-one-dimensional physics—against strictly one-dimensional techniques. Such an endeavor could involve modeling the system as a series of coupled wires using tensor network methods~\cite{schollwock2005densitymatrix, szasz2021phase} or utilizing auxiliary-field or stochastic series expansion quantum Monte Carlo simulations, which remain sign-problem-free in the limit of vanishing inter-wire hopping or at half filling~\cite{li2019signproblemfree, calugaru2026mixed, vasiliou2026hidden}.

\medskip

\prlparagraph{Acknowledgments}%
L.K., S.V.S., and R.T.~acknowledge support by the Deutsche Forschungsgemeinschaft (DFG, German Research Foundation) through Project-ID 258499086 -- SFB 1170 and through the Würzburg-Dresden Cluster of Excellence on Complexity and Topology in Quantum Matter -- ctd.qmat Project-ID 390858490 -- EXC 2147.
Additionally S.V.S.~was supported by the Hanns-Seidel Foundation and the Three Physicists Philanthropic Trust.
A.F.~is funded by the Deutsche Forschungsgemeinschaft (DFG, German
Research Foundation) -- 572935092.
B.A.B.~and H.H.~were supported by the Gordon and Betty Moore Foundation through Grant No.~GBMF8685 towards the Princeton theory program, the Gordon and Betty Moore Foundation’s EPiQS Initiative (Grant No.~GBMF11070), the Global Collaborative Network Grant at Princeton University, the Simons Investigator Grant No.~404513, the Princeton Catalysis Initiative, the NSF-MERSEC (Grant No.~MERSEC DMR 2011750), the Simons Collaboration on New Frontiers in Superconductivity (Grant No.~SFI-MPS-NFS-00006741-01), and the Schmidt Fund at the Princeton University.
The Flatiron Institute is a division of the Simons Foundation.
H.H.~acknowledges support from the European Research Council (ERC) under the European Union’s Horizon 2020 research and innovation program (Grant Agreement No.~101020833).

\let\oldaddcontentsline\addcontentsline
\renewcommand{\addcontentsline}[3]{}
\bibliography{FRG_cleaned.bib}
\let\addcontentsline\oldaddcontentsline

\supplement{Supplementary Information:\\\titlePaper}
\tableofcontents

\section{Interacting Wannier Model}
\label{app:wannier-interaction}
The hopping parameters $t_{\bv R,\eta}$ of the kinetic Hamiltonian \cref{eq:hkin} are obtained from Wannier projection of the continuum Hamiltonian in momentum space, i.e.,
\begin{equation}
    t_{\bv k,\eta} = \sum_{\bv G,\bv G',l,l'} \big(\mathcal W_{\bv k,\eta}^{\bv G,l}\big)^* H_{\bv k,\eta}^{\bv G,\bv G',l,l'} \mathcal W_{\bv k,\eta}^{\bv G',l'} \,.
\end{equation}
Here, $\mathcal W_{\bv k,\eta}^{\bv G,l}$ labels the Wannier function of valley $\eta$ at moir\'e lattice vector $\bv G$ and layer $l$ for the Bloch momentum $\bv k$ and $H^{\bv G,\bv G',l,l'}_{\bv k,\eta}$ is the continuum Hamiltonian (see Supplemental Material of Ref.~\cite{li2025emergent} for details and \cref{tab:wannparam} for parameters obtained from \emph{ab initio} simulations).

\begin{table}
    \centering
    \caption{Hopping parameters for the Wannier models of AB-stacked \ch{tSnSe2} for various twist angles $\theta$. All values given in $\mathrm{meV}$. The definition of hopping parameters follows \cref{fig:model}(a). Throughout this work, we use the parametrization for $\theta=6.01^\circ$ (bold).}
    \label{tab:wannparam}
    \begin{ruledtabular}
    \begin{tabular}{ccccccc}
         $\theta$ & $t_{\parallel1}$ & $t_{\parallel2}$ & $t_{\perp1}$ & $t_{\setminus1}$ & $t_{\setminus2}$ & $t_{\setminus3}$ \\[0.3em]\hline
$3.89^\circ$ & $0.6448$ & $0.0250$ & $0.0698$ & $\phantom{-}1.2973$ & $0.0619$ & $0.0048$ \\
$4.41^\circ$ & $1.0530$ & $0.0538$ & $0.1590$ & $\phantom{-}1.3614$ & $0.0987$ & $0.0107$ \\
$5.09^\circ$ & $2.6149$ & $0.2191$ & $0.3894$ & $\phantom{-}1.2520$ & $0.1948$ & $0.0384$ \\
$\bf{}6.01^\circ$ & $\bf{}4.3620$ & $\bf{}0.5079$ & $\bf{}0.9848$ & $\bf{}\phantom{-}0.8446$ & $\bf{}0.2584$ & $\bf{}0.0768$ \\
$7.34^\circ$ & $9.8779$ & $1.6289$ & $4.0917$ & $-0.2600$ & $0.3111$ & $0.1586$ \\
    \end{tabular}
    \end{ruledtabular}
\end{table}

As shown in \cref{tab:wannparam}, the hopping parameters exhibit a strong
dependence on the twist angle $\theta$, driving a significant dimensional
crossover. To quantify this effect, we define the kinetic
anisotropy
$\alpha = | t_{\setminus 1}/t_{\parallel 1} |$,
where $t_{\parallel 1}$ denotes the hopping along the primary direction and
$t_{\setminus 1}$ denotes the hopping along the inequivalent nearest-neighbor
direction. The resulting anisotropies for both AA and AB stacking are summarized
in \cref{tab:anisotropy}.

\begin{table}
\centering
\caption{Kinetic anisotropy for AA- and AB-stacked SnSe2. We define
$\alpha = |t_{\setminus 1}/t_{\parallel 1}|$. Hopping amplitudes are in meV.}
\label{tab:anisotropy}
\begin{ruledtabular}
\begin{tabular}{ccccccc}
$\theta$ &
\multicolumn{3}{c}{AA stacking} &
\multicolumn{3}{c}{AB stacking} \\
\cline{2-4}\cline{5-7}
&
$t_{\parallel 1}$ & $t_{\setminus 1}$ & $\alpha$ &
$t_{\parallel 1}$ & $t_{\setminus 1}$ & $\alpha$ \\
\hline
$3.89^\circ$ & $1.399$ & $-0.231$ & $0.165$ & $0.645$ & $\phantom{-}1.297$ & $2.012$ \\
$4.41^\circ$ & $2.217$ & $-0.133$ & $0.060$ & $1.053$ & $\phantom{-}1.361$ & $1.293$ \\
$5.09^\circ$ & $4.220$ & $-0.105$ & $0.025$ & $2.615$ & $\phantom{-}1.252$ & $0.479$ \\
$6.01^\circ$ & $5.886$ & $\phantom{-}0.063$ & $0.011$ & $4.362$ & $\phantom{-}0.845$ & $0.194$ \\
$7.34^\circ$ & $11.350$ & $\phantom{-}0.467$ & $0.041$ & $9.878$ & $-0.260$ & $0.026$ \\
\end{tabular}
\end{ruledtabular}
\end{table}

\begin{figure}
    \includegraphics{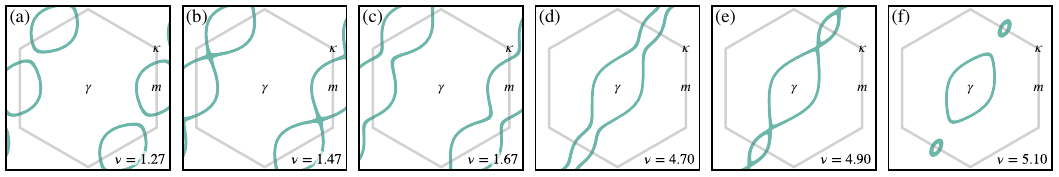}
    \caption{\label{fig:supp-fs}Fermi surfaces of valley $\eta=\mathrm{C}$ at fillings close to the lower (a-c) and upper (d-f) Van Hove filling. The pockets~(a) and lines~(c) emerging from the Van Hove singularity around $\nu\approx 1.47$~(b) exhibit significantly more two-dimensional character than the pockets~(f) and lines~(d) that emerge from the Van Hove singularity at $\nu\approx 4.90$~(e).}%
\end{figure}

For AB stacking, $\alpha$ decreases strongly with increasing
twist angle, from $\alpha\simeq 2.01$ at $3.89^\circ$ to
$\alpha\simeq 0.026$ at $7.34^\circ$. This reflects a change in the dominant
hopping direction: at small twist angles $t_{\setminus 1}$ is larger than
$t_{\parallel 1}$, whereas at larger twist angles $t_{\parallel 1}$ becomes
dominant. The kinetic model is therefore most anisotropic at both ends of this
angle range, while intermediate angles, especially around $4.41^\circ$--$5.09^\circ$,
are closer to a two-dimensional hopping network. At our focus angle
$\theta=6.01^\circ$, the anisotropy is already sizable,
$\alpha\simeq 0.19\approx 1/5$, comparable in magnitude to AA-stacked \ch{SnSe2} at
$\theta=3.89^\circ$, where $\alpha\simeq 0.17$. We display Fermi surfaces of valley $\eta=\mathrm{C}$ in \cref{fig:supp-fs}.

This quasi-one-dimensional feature is the fundamental driver for the banded
Fermi surfaces and the nesting that promotes high-energy antiferromagnetic
fluctuations discussed in the main text.  Furthermore, the tight-binding model
constructed strictly with the truncated parameters in \cref{tab:wannparam}
accurately reproduces the low-energy band dispersion of the full continuum
model, validating our kinetic truncation.

\begin{SCfigure}
    \centering
    \includegraphics{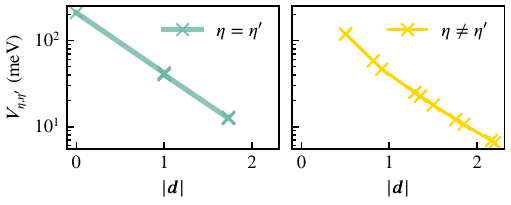}
    \caption{\label{fig:suppvert}%
        Bare coulomb repulsion in the Wannier basis for $\epsilon=12$ split into intravalley (left) and intervalley (right) components and plotted as a function of distance in units of moir\'e lattice vectors ($|\bv d|$, see \cref{tab:vertices}).
        The slight discontinuity at $|\bv d|=1$ in the intravalley case corresponds to marginally different values for identical distances, which is allowed because of the Wannier orbitals' low symmetry.
    }
\end{SCfigure}

\begin{table}
\begin{minipage}[c]{0.7\textwidth}
\begin{ruledtabular}
\begin{tabular}{cccc}
$|\bv d|$ & $V_\mathrm{\eta=\eta'}/\mathrm{meV}$ &
$|\bv d|$ & $V_\mathrm{\eta\neq\eta'}/\mathrm{meV}$ \\\hline
0                & 210.995 & 0.502303 & 119.259 \\
$1_\parallel$    & 41.9850 & 0.817985 & 58.2029 \\
$1_\setminus$    & 40.0978 & 0.914066 & 46.7610 \\
$1.73205_\setminus$ & 12.8290 &  1.29194 & 24.9636 \\
$1.73205_\perp$  & 12.4946 &  1.35481 & 22.6628 \\
                 &         &  1.50077 & 17.8388 \\
                 &         &  1.75667 & 12.1053 \\
                 &         &  1.84898 & 10.6460 \\
                 &         &  2.16081 & 6.89255 \\
                 &         &  2.19898 & 6.53657 \\
\end{tabular}
\end{ruledtabular}
\end{minipage}
\hfill
\begin{minipage}[c]{0.27\textwidth}
\caption{Values of the density-density interaction matrix elements in the Wannier basis for intravalley (left two columns, $V_{\eta=\eta'}$) and intervalley (right two columns, $V_{\eta\neq\eta'}$) interactions of $\theta=6.01^\circ$ AB-stacked \ch{SnSe2}; other angles given in the git repository~\cite{klebl2026mtwist}. The first (third) column denotes the distance of the corresponding density-density interaction in units of moir\'e lattice vectors ($|\bv d|$). The subscripts correspond to parallel/skew/perpendicular directions of equivalent distances, cf.~\cref{fig:model}.}
\label{tab:vertices}
\end{minipage}
\end{table}

The Coulomb repulsion is modeled using a dual-gated screened potential $V(\bv q) = 2\pi e^2\tanh(\xi|\bv q|/2)/(\epsilon|\bv q|)$. The screening length $\xi = 10~\mathrm{nm}$ reflects the typical distance to the experimental gates. The dielectric constant $\epsilon = 12$ serves as a baseline derived from rigorous density functional perturbation theory (DFPT) calculations, which carefully account for the intrinsic 2D dielectric response of the \ch{SnSe2} bilayer and the environmental screening from \ch{hBN} encapsulation~\cite{li2025emergent}. This interaction is projected to the Wannier basis following the procedure formulated in momentum space in Ref.~\cite{li2025emergent}. The resulting density-density interaction components as a function of real space distance are shown in \cref{fig:suppvert} (for $\theta=6.01^\circ$). We include all density-density interaction components whose distance is less than $2.2$ moir\'e lattice vectors.

While the rigorous Wannier projection theoretically generates non-density-density terms, such as spin-spin couplings and density-dependent hoppings, numerical evaluations demonstrate that these terms are energetically marginal (on the order of $\sim 2~\mathrm{meV}$)~\cite{li2025emergent}. In stark contrast, at $\theta=6.01^\circ$, the dominant intravalley on-site Hubbard $U$ is approximately $105.5~\mathrm{meV}$, and the nearest-neighbor intervalley density-density interaction $V_1$ is approximately $59.6~\mathrm{meV}$. We therefore expect that restricting the interaction Hamiltonian strictly to density-density terms is a robust approximation.

\section{Truncated unity FRG (TUFRG)}
\label{app:tufrg}
In the static approximation to the functional renormalization group, the central object is the scale-dependent four-point vertex
\begin{equation}
    \Gamma^\Lambda_{o_1o_2o_3o_4}(\bv k_1,\bv k_2,\bv k_3)\,,
\end{equation}
which describes the effective interaction between two incoming and two outgoing fermions. Here \(\bv k_i\) label momenta, and \(o_i\) orbitals or sublattices. During the FRG flow, this vertex develops a complicated dependence on momenta and orbital indices, which renders storing it on a full three-momentum/four-orbital grid numerically very expensive.

The truncated-unity FRG (TUFRG) is a way to compress this momentum dependence. The basic idea is to resolve the dependence on the two ``relative'' momenta in a finite basis of lattice form factors. These form factors may be viewed as real-space bonds connecting two orbitals in possibly different unit cells. A bond is labeled by
\begin{equation}
    b \equiv (o_i,o_j,\bv R) \,,
\end{equation}
where the vector $\bv r_b = \bv o_i-\bv o_j+\bv R$ connects orbital \(o_j\) in unit cell \(\bv R\) to orbital \(o_i\) in the reference cell. In momentum space, this bond corresponds to a phase factor, schematically
\begin{equation}
    b(o_i,o_j,\bv k) \sim e^{i\bv k\cdot \bv r_b} \,.
    \label{eq:bondbasis}
\end{equation}
The full set of all such bonds would form a complete basis for the relative-momentum dependence of the vertex. In practice, TUFRG keeps only bonds shorter than some cutoff distance,
\begin{equation}
    |\bv r_b| < r_{\max} \,.
\end{equation}
This is the ``truncation'' in truncated unity. Physically, it assumes that the important structure of the interaction vertex can be represented by short- and intermediate-range form-factors. Increasing \(r_{\max}\) systematically improves the approximation.

In order to truncate it, the vertex is decomposed into the three two-particle reducible channels:
\begin{equation}
    \mathcal X \in \{\mathcal P,\mathcal C,\mathcal D\} \,,
\end{equation}
where \(\mathcal P\) is the particle-particle (superconducting) channel, \(\mathcal C\) is the crossed particle-hole (magnetic) channel, and \(\mathcal D\) is the direct particle-hole (charge) channel. Each channel is naturally described by one bosonic transfer momentum \(\bv q\) and two fermionic, relative momenta. TUFRG expands the two relative-momentum dependences in the bond basis [cf.~\cref{eq:bondbasis}], leaving only the transfer momentum \(\bv q\) explicit. Thus, instead of storing a complicated function of three momenta, one stores matrices
\begin{equation}
    \mathcal X^\Lambda_{bb'}(\bv q) \,,
\end{equation}
where \(b\) and \(b'\) are \emph{truncated} bond (or form-factor) indices.

Schematically, the channel projection has the form
\begin{equation}
    \mathcal X[Y]_{bb'}(\bv q)
    =
    \sum_{\bv k,\bv k'}
    b(\bv k)\,
    Y(\bv q,\bv k,\bv k')\,
    b'^*(\bv k') \,,
\end{equation}
with different choices of \(\bv q,\bv k,\bv k'\) for the three channels. More explicitly, for a general four-point vertex \(Y_{o_1\ldots o_4}\), the projections are
\begin{align}
    \label{eq:frgproj-p}
    \mathcal P[Y]_{bb'}(\bv q)
    &{}= \sum_{\bv k_1,\bv k_3,o_1,\dots,o_4}
    b(o_1,o_2,\bv k_1)
    b'^*(o_3,o_4,\bv k_3)
    Y_{o_1\ldots o_4}
    (\bv k_1,\bv q-\bv k_1,\bv k_3) \,,
    \\
    \label{eq:frgproj-c}
    \mathcal C[Y]_{bb'}(\bv q)
    &{}= \sum_{\bv k_1,\bv k_3,o_1,\dots,o_4}
    b(o_1,o_4,\bv k_1)
    b'^*(o_3,o_2,\bv k_3)
    Y_{o_1\ldots o_4}
    (\bv k_1,\bv k_1-\bv q,\bv k_3) \,,
    \\
    \label{eq:frgproj-d}
    \mathcal D[Y]_{bb'}(\bv q)
    &{}= \sum_{\bv k_1,\bv k_4,o_1,\dots,o_4}
    b(o_1,o_3,\bv k_1)
    b'^*(o_4,o_2,\bv k_4)
    Y_{o_1\ldots o_4}
    (\bv k_1,\bv k_4-\bv q,\bv k_1-\bv q) \,.
\end{align}
The notation \(b(o_i,o_j,\bv k)\) denotes the Fourier transform of the real-space bond basis function, cf.~\cref{eq:bondbasis}.

The advantage of this representation becomes visible in the FRG flow equations. The one-loop flow contains bubble diagrams in the particle-particle and particle-hole channels. After projection onto the bond basis, these loop contributions become matrix products over the bond indices \(b,b'\) at fixed transfer momentum \(\bv q\). Thus, the complicated momentum convolutions of the original FRG equations are replaced by numerically efficient operations on matrices \(\mathcal X_{bb'}(\bv q)\) with truncated dimension $b$.

The loop kernels, or differentiated susceptibilities, must therefore also be represented in the same bond basis. For the particle-particle and particle-hole bubbles one obtains
\begin{equation}
\label{eq:suscb}
    \dot\chi^{P/H}_{\Lambda;bb'}(\bv q)
    =
    \sum_{\bv k,o_1,\ldots,o_4}
    b(o_1,o_2,\bv k)
    b'^*(o_3,o_4,\bv k)
    \dot\chi^{P/H}_{\Lambda;o_1\ldots o_4}(\bv q,\bv k) \,,
\end{equation}
with
\begin{equation}
\label{eq:susck}
\dot\chi^{P/H}_{\Lambda;o_1\ldots o_4}(\bv q,\bv k)
=
\frac{1}{2\pi}
\big[
G^0_{o_1o_3}(\pm i\Lambda,\bv k)
G^0_{o_2o_4}(-i\Lambda,\pm\bv q\mp\bv k)
+
G^0_{o_1o_3}(\mp i\Lambda,\bv k)
G^0_{o_2o_4}(+i\Lambda,\pm\bv q\mp\bv k)
\big] \,.
\end{equation}
The upper and lower signs correspond to the particle-particle and particle-hole bubbles, respectively. Once these bubble matrices are computed, the \(SU(2)\)-symmetric FRG flow equations for the three channels can be written as matrix equations in the bond basis. In summary, the TUFRG flow equations read
\begin{align}
    \label{eq:frg-P}
    \dot{\mathcal P}^\Lambda &{}= \mathcal P[\Gamma^{(4)}_\Lambda] \circ \dot{\chi}^P_\Lambda \circ \mathcal P[\Gamma^{(4)}_\Lambda] \,, \\
    \label{eq:frg-C}
    \dot{\mathcal C}^\Lambda &{}= \mathcal C[\Gamma^{(4)}_\Lambda] \circ \dot{\chi}^H_\Lambda \circ \mathcal C[\Gamma^{(4)}_\Lambda] \,, \\
    \label{eq:frg-D}
    \dot{\mathcal D}^\Lambda &{}= -2
        \bigg(\mathcal D[\Gamma^{(4)}_\Lambda] - \frac12 \mathcal C[\Gamma^{(4)}_\Lambda]\bigg)
        \circ
        \dot{\chi}^H_\Lambda
        \circ
        \bigg(\mathcal D[\Gamma^{(4)}_\Lambda] - \frac12 \mathcal C[\Gamma^{(4)}_\Lambda]\bigg)
        + \frac{\dot{\mathcal C}^\Lambda}2 \,,
    \\
    \label{eq:frg-V}
    \Gamma^{(4)}_\Lambda &{}= \sum_{\mathcal X\in\{\mathcal P,\mathcal C,\mathcal D\}} \mathcal X^{-1}[\mathcal X^\Lambda] \,,
\end{align}
where $\mathcal X^{(-1)}[Y]$ denotes the (inverse) projection of the vertex $Y$ into channel $\mathcal X$ [cf.~\cref{eq:frgproj-p,eq:frgproj-c,eq:frgproj-d}], ``$\circ$'' represents a matrix product over all internal quantum numbers (i.e., bonds) and a regular product over diagonal quantities (i.e., transfer momentum $\bv q$), and $\dot{\chi}^{P/H}_\Lambda$ is the at-scale particle-particle/particle-hole susceptibility [cf.~\cref{eq:suscb,eq:susck}]. Note that we follow the standard procedure of fixing the spin indices of the vertex functions to the $\Gamma^{(4)}_{\uparrow\downarrow;\downarrow\uparrow}$ sector. The such obtained TUFRG flow equations are implemented in the divERGe library~\cite{profe2024diverge, profe2024codebase}, which we use to run this work's simulations. The code is accessible at Ref.~\cite{klebl2026mtwist}, details are given in Ref.~\cite{frg-details}.

\subsection{Effective Hubbard Interaction}
\begin{SCfigure}
    \centering
    \includegraphics{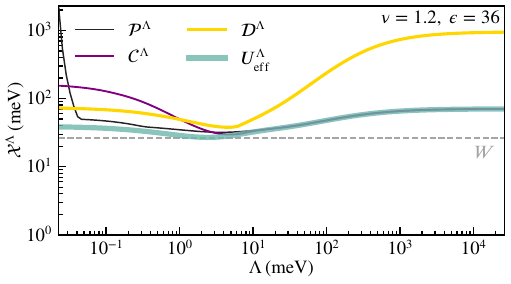}
    \caption{\label{fig:suppflow}%
        FRG flow visualizing the leading running couplings in each of the diagrammatic channels ($\mathcal P^\Lambda$, $\mathcal C^\Lambda$, $\mathcal D^\Lambda$) as well as the running Hubbard interaction $U_\mathrm{eff}^\Lambda$, cf.~\cref{eq:effective_hubbard}. The bandwidth $W$ is indicated as dashed gray line. The couplings shown are for $\nu=1.2$ and $\epsilon=36$. Note that $\mathrm{min}_\Lambda (U_\mathrm{eff}^\Lambda)\approx W$.
    }
\end{SCfigure}

As the model includes multiple long-ranged as well as inter-orbital interaction components, there is significant screening from the $\mathcal D$-channel at large RG scales. The effective Hubbard interaction,
\begin{equation}
    \label{eq:effective_hubbard}
    U^\Lambda_\mathrm{eff} = \sum_{\bv k_1,\bv k_2,\bv k_3} [\Gamma^{(4)}_\Lambda]_{\eta\eta\eta\eta}(\bv k_1,\bv k_2,\bv k_3) \,,
\end{equation}
is strongly affected by this, as shown in \cref{fig:suppflow} along with the maxima of each of the channels.

\FloatBarrier

\begin{SCfigure}
    \centering
    \includegraphics{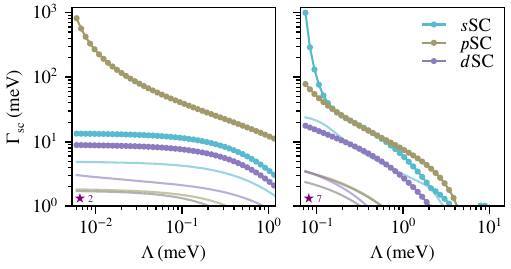}
    \caption{FRG flow of the pairing vertex $\Gamma_\mathrm{sc}$ for the second (left) and seventh (right) purple star in \cref{fig:frg-phases}. The leading and subleading eigenvalues are colored according to their irrep, i.e., blue ($s$SC), olive ($p$SC), and purple ($d$SC).}
    \label{fig:star-flows}
\end{SCfigure}

\subsection{Banded Structure of the AFM Vertex}
\label{app:banded-vertex}
\begin{SCfigure}
    \centering
    \includegraphics{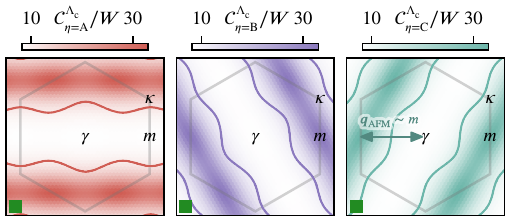}
    \caption{\label{fig:supptube}%
        Intravalley $\mathcal C$-channel vertex from FRG at the critical scale $\Lambda_\mathrm{c}$ for half filling $\nu=3$ and $\epsilon=12$ (green square in \cref{fig:frg-phases}). $W$ denotes the bandwidth. Each panel displays the vertex $\mathcal C_\eta^{\Lambda_\mathrm c}$ for one of the valleys $\eta$ (A-C from left to right; coloring follows \cref{fig:model}). The Fermi surface is overlaid as line of the respective valley's color, clearly displaying both quantities' one-dimensional character.
        We indicate a representative AFM fluctuation vector $\bv q_\mathrm{AFM}$ in the rightmost panel.
    }
\end{SCfigure}
To demonstrate the quasi one-dimensional nature of the AFM $\mathcal C$-channel vertex, \cref{fig:supptube} displays all three intravalley components as well as the respective Fermi surfaces for the point in parameter space represented by the green square in \cref{fig:frg-phases}.

\section{Intraorbital bilinear approximation (IOBI)}
\label{app:iobi}
TUFRG with zero maximal bond distance ($r_\mathrm{max}=0$, cf.~\cref{app:tufrg}) is also known as intraorbital bilinear approximation (IOBI)~\cite{honerkamp2018efficient, klebl2020functional}. It differs only slightly from individual ladder resummations (i.e., RPA) in all channels; the on-site components of the interaction are renormalized simultaneously as they can be written in any of the channels.

Notably, screening from the $\mathcal P$- and $\mathcal D$-channels is incorporated in parallel with Stoner-like behavior in the $\mathcal C$-channel through interchannel feedback. For \ch{tSnSe2} in particular, this means that the magnetic instability is pushed to significantly smaller $\Lambda_\mathrm c$.
Similar to an RPA treatment of superconducting instabilities, we use the renormalized vertex in the exchange channel $\mathcal C_{\eta,\eta'}^{\Lambda_\mathrm c}(\bv q)$ as input to the linearized gap \cref{eq:lingap}, which demonstrates that spin fluctuations act as pairing glue, see \cref{fig:pwave}. In order to obtain vertices of reasonable absolute value, we terminate the IOBI flow when any channel reaches an eigenvalue larger than $3.5W$ (smaller than our termination criterion for FRG), with $W\approx 26.4\,\mathrm{meV}$ the bandwidth. All other parameters are equivalent to those used in the FRG simulations, see Ref.~\cite{frg-details}.

\subsection{Linearized Superconducting Gap Equation}
\label{app:lingap}
\begin{SCfigure}
    \centering
    \includegraphics{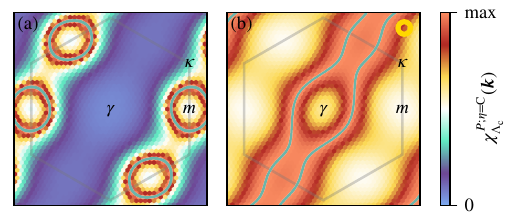}
    \caption{\label{fig:suppchi}%
        Particle-particle susceptibility evaluated at the critical (IOBI) scale for $\nu=0.72$, $\epsilon=33$~(a) and $\nu=4.44$, $\epsilon=16$~(b, golden circle in \cref{fig:frg-phases} of the main text) that correspond to $p$SC~(a) and $s$SC~(b), respectively. The $p$SC case~(a) has circular features at the Fermi level, whereas the $s$SC case~(b) is mostly one-dimensional. In both panels, the Fermi surface is indicated as green line and we plot only the $\eta=\mathrm C$ valley. The significantly lower critical scale $\Lambda_\mathrm c$ in panel~(a) makes the Fermi surface features that $\chi^P$ encodes way sharper than the ones of panel~(b).
    }
\end{SCfigure}
As noted in the previous appendix, we set up the linearized superconducting gap \cref{eq:lingap} from the renormalized vertex in the $\mathcal C$-channel obtained from the IOBI simulation. Note that we only use the intravalley component as all the phases we find within FRG are intravalley as well.
The intravalley particle-particle susceptibility is given by
\begin{equation}
    \chi^P_{\eta,\Lambda_\mathrm c}(\bv k) = \frac{f\big(-\pi\epsilon_\eta(\bv k)/\Lambda_\mathrm c\big)-f\big(\pi\epsilon_\eta(-\bv k)/\Lambda_\mathrm c\big)}{\epsilon_\eta(\bv k)+\epsilon_\eta(-\bv k)} \,,
    \label{eq:particle-particle-loop}
\end{equation}
where $f(x)=(1+e^x)^{-1}$ is the Fermi function, and we use $\Lambda_\mathrm c/\pi$ as temperature---motivated by the value of the lowest Matsubara frequency.
We evaluate \cref{eq:particle-particle-loop} on the same $36\times36$ momentum mesh as the vertex, see \cref{fig:suppchi}.

\end{document}